# RADAR-BASED RAINDROP SIZE DISTRIBUTION PREDICTION: COMPARING ANALYTICAL, NEURAL NETWORK, AND DECISION TREE APPROACHES

R. J. Humphreys

**ABSTRACT**: Reliable estimation of the raindrop size distribution (RSD) is important for applications including Quantitative Precipitation Estimation (QPE), soil erosion modelling, and wind turbine blade erosion. While in situ instruments such as disdrometers provide detailed RSD measurements, they are spatially limited, motivating the use of polarimetric radar for remote retrieval of rain microphysical properties. This study presents a comparative evaluation of analytical and machine-learning approaches for retrieving RSD parameters from polarimetric radar observables. One-minute OTT Parsivel2 disdrometer measurements collected between September 2020 and May 2022 at Sheepdrove Farm, UK, were quality-controlled using collocated weighing and tipping-bucket rain gauges. Measured RSDs were fitted to a normalised three-parameter gamma distribution, from which a range of polarimetric radar variables were analytically simulated. Analytical retrievals, neural networks, and decision tree models were then trained to estimate the gamma distribution parameters across multiple radar feature sets and model architectures. To assess robustness and equifinality, each model configuration was trained 100 times using random 70/30 train-test splits, yielding approximately 17,000 trained models in total. Machine-learning approaches generally outperform analytical methods; however, no single model class or architecture is uniformly optimal. Model performance depends strongly on both the target RSD parameter and the available radar observables, with decision trees showing particular robustness in reduced-feature regimes. These results highlight the importance of aligning retrieval model structure with operational data constraints rather than adopting a single universal approach.

**KEYWORDS**: Raindrop Size Distribution; Quantitative Precipitation Estimation; Radar; Machine Learning

## 1. INTRODUCTION

Obtaining reliable estimates of the Raindrop Size Distribution (RSD) from precipitation events is highly important in various real-world problems, such as soil erosion, wind turbine blade erosion, and quantitative precipitation estimation. Traditional methods for obtaining the RSD involve deploying expensive ground-based empirical measuring equipment (e.g., disdrometers or optical devices). However, these measurements only provide estimates at point locations: they fail to capture the spatial variability of the RSD. Instead, researchers use polarimetric radar. Polarimetric radar obtains orthogonally polarised reflectivity measurements that may be used to remotely estimate the microphysical properties of rain, and thus, the rainfall rate. The ability of this technology to provide a cheaper and more convenient method of estimating the RSD has been explored in various studies, but more work is needed to robustly evaluate and compare these models.

This study leverages OTT Parsivel$^2$ disdrometer data that was collected at one-minute intervals between September 2020 and May 2022 from Sheepdrove Farm, Swindon. In addition to the quantity of raindrops of certain diameters passing through the sensor, the disdrometer data included rainfall rate and hydrometeor type. The hydrometeor type was used to discard all non-liquid precipitation (e.g., hail). Data from three further rain gauges was also available for the same site: a weighing gauge, and two tipping bucket rain gauges (TBRGs). Consequently, it was possible to assess the validity of the disdrometer data by comparing its results against the data provided by the other rain gauges.

Using the disdrometer data, measured RSDs were obtained and fitted to the normalised three-parameter gamma model of the RSD using established parameter retrieval techniques. The modelled RSD was then used to analytically simulate an extensive range of radar parameters including the reflectivity and differential reflectivity. Various approaches common in the literature, including analytical, neural network, and decision tree models were then trained to estimate the parameters of the normalised gamma model. For the neural network and decision tree approaches, the model's structure was varied, enabling a comparison of the skilfulness of different architectures. Further, the models were provided with a wide range of radar products to account for the varying availability and reliability of radar products in real world applications. Because the models are data centric, each model was trained 100 times from a random 70% of the simulated radar parameters, and the remaining 30% of the data was reserved for testing the models. In total, 17000 unique models were produced, thus enabling broad and rigorous insights into the efficacy of different RSD retrieval models.

The results showed that machine learning approaches are more accurate than the analytical methods at obtaining the RSD parameters. However, no single machine learning model showed consistently more skill at predicting the parameters of the RSD. Instead, the way that a model is structured should be contingent on the range of radar data obtainable and the particular parameter the model is finding. For instance, decision trees with 36 leaf nodes proved to be the most capable model at accurately obtaining the median volume drop diameter given only radar reflectivity and differential reflectivity measurements.

**STATEMENT OF ORIGINALITY**: This paper trained an extensive array of different RSD retrieval models based on those proposed in the literature. The input data was randomised 100 times, enabling a more rigorous analysis of the model's predictive utility than previous studies. Further, given the concept of equifinality, a wider range of radar parameters were mobilised to calibrate the machine learning (ML) models. This accounted for the varying availability and uncertainty of radar products in real world data. Furthermore, a far broader range of ML structures were tested to evaluate the impact of model architecture on performance. Consequently, the paper facilitates a broad and deep comparison of how different model types, inputs parameters, and ML structures affect RSD retrieval accuracy.

## 2. LITERATURE REVIEW

Accurate estimation of the raindrop size distribution (RSD) is pivotally important across various real-world problems. For instance, understanding the RSD enables more accurate insights into the kinetic energy of rainfall. This insight is valuable when predicting the level of leading-edge erosion of wind turbine blades. Turbine blade erosion is caused by raindrops colliding with the blades and causing tiny defects, which increases the roughness of the blades, decreasing their aerodynamic performance, and thus decreasing the overall energy output. The level of erosion is linked to the kinetic energy of the incident rain (Tilg et al., 2022). Similarly, Caracciolo et al. (2012) stated that soil erosion is also linked to the kinetic energy of the rainfall and that the RSD 'is well correlated to the measured soil loss'. Thus, models including those proposed by Dai et al. (2020) demonstrate that accurate soil erosion models can be developed using high-resolution RSDs.

Understanding the RSD is similarly crucial in quantitative precipitation estimation (QPE) because it provides insights into the microphysical properties of rain, which can be used to improve the accuracy of radar rainfall measurements and predictions (Tokay and Short, 1996; Ulbrich, 1983). However, radar systems cannot directly measure properties such as the rainfall rate, instead, they obtain values for radar parameters which can be subsequently used to estimate rainfall properties. For instance, weather radar systems can be used to estimate the size, shape, and concentration of precipitation by measuring radar parameters such as the reflectivity and the differential reflectivity. For example, the reflectivity is related to the $6^{th}$ moment of the RSD (Seliga and Bringi, 1976). This relationship is especially critical for QPE because the rainfall rate is the $3.67^{th}$ moment of the RSD (Zhang et al., 2001;Cluckie and Rico-Ramirez, 2004). Consequently, enhancing our understanding, modelling, and prediction of the RSD can significantly improve radar-derived rainfall rate estimations.

### 2.1 Measuring the Raindrop Size Distribution

The RSD is a statistical representation of the range and frequency of raindrop sizes during a precipitation event. Throughout the twentieth century, various methods were developed to measure the RSD. For instance, the stain method was a frequently used measurement technique. By exposing a chemically treated paper to rain for a short period of time, experimenters were able to record the stains left behind (Kathiravelu et al., 2016). An example of this comes from Marshall and Palmer (1948), who used two dyed filter papers to record the drop sizes and thus fit an equation to the RSD. Another common RSD measurement technique is the flour pellet method. Here, binding agent and flour is briefly exposed to rain, wetting the flour, and forming hard pellets. This method is often used in soil erosion research (Kathiravelu et al., 2016).

More recently, various types of disdrometers have been developed to measure the RSD. One such example is the Joss-Waldvogel (JW) displacement disdrometer. The JW disdrometer has been used in various studies: for instance, Montopoli et al. (2008) employed JW disdrometer data to analyse and model the temporal evolution of normalised gamma RSDs. Similarly, Tokay et al. (2003) employed collocated JW disdrometers to study the range dependency of radar calibration and rainfall verification in tropical storms.

Other studies have used video disdrometers to measure the RSD. For example, Adirosi et al. (2014) used a video disdrometer to analyse the accuracy of the three-parameter gamma RSD against the natural RSD. Further, Bringi et al. (2002) used video disdrometer data to propose a method for estimating the parameters of a gamma raindrop size distribution model from radar measurements.

The present research will use an OTT Parsivel$^2$ optical disdrometer. Tokay et al. (2014) demonstrated that the Parsivel$^2$ exhibited stronger agreement with reference rain gauges than both the Parsivel$^1$ and the aforementioned JW disdrometer. The Parsivel$^2$ has been used in studies including Tang et al. (2014) which compared the RSD retrieved from polarimetric radar to the measurements obtained from the Parsivel$^2$. Here, researchers further confirmed that there are large rainfall microphysical variations across different climatic regions.

Thus far, various ground-based empirical measurement techniques for obtaining the RSD and the rainfall rate have been introduced. However, despite their utility, they are unable to capture the spatial variability of rain over large areas (Bringi et al., 2002; Bringi and Chandrasekar, 2001). To overcome this limitation, radar technology has been developed as a remote sensing technique that can provide detailed information about precipitation at a distance. There are two principal types of radar systems: single-polarimetric and dual-polarimetric. Both systems transmit and receive pulses of radio waves, allowing them to measure phenomena including the radar reflectivity and Doppler velocity (Sokol et al., 2021).

The seminal study by Seliga and Bringi (1976) demonstrated the potential of orthogonally polarised radar for estimating rain rate. This research led to the development of dual-polarised radar systems that offer several advantages over the earlier single-polarised radar systems (Sokol et al., 2021). For instance, dual-polarised radar provided additional radar products, including: the horizontal or vertical reflectivity ($Z_{h,v}$), the differential reflectivity ($Z_{dr}$), the specific differential phase shift ($K_{dp}$), the specific attenuation ($A_H$), the correlation coefficient ($\rho_{hv}$), and the linear depolarisation ratio ($LDR_{vh}$).

This technology has been applied to both ground-based and active satellite rainfall sensors. Active satellite rainfall sensors are radar instruments that are mounted on satellites. They are crucial for rainfall rate prediction as they provide extensive global coverage, including over vast, inaccessible areas, such as oceans and remote regions. For instance, Bringi et al. (2002) used the Tropical Rainfall Measuring Mission (TRMM) satellite to obtain data radar products during a squall-line event over Brazil.

## 2.2 Modelling and Retrieval of Raindrop Size Distribution

Rain's microstructure is defined by its RSD. Thus, significant effort has been expended attempting to develop a parameterised model that accurately describes the shape of the RSD. For instance, one of the first attempts was proposed by Marshall and Palmer (1948) who suggested that the RSD follows a simple two parameter exponential distribution. Further developments from Seliga and Bringi (1976) demonstrated that the dual-polarimetric differential reflectivity parameter ($Z_{dr}$) is related to droplet size. Therefore, in their seminal paper, the researchers were able to show that polarimetric radar can be used to estimate the median volume diameter ($D_0$) described by the exponential model of the RSD. However, later research conducted by Ulbrich (1983) showed that Marshall and Palmer's (1948) exponential distribution was not sufficiently capable of describing the RSD; instead, he proposed that the RSD is more truthfully modelled using the three parameter ($N_w$, $D_0$, $\mu$) gamma model. Building upon this work, Testud et al. (2001) introduced the (three parameter) normalised gamma distribution to better represent the natural RSD found across the globe. Finally, by exploring the RSD from diverse climatic regimes (including: continental, maritime, and equatorial regions), Bringi et al. (2003) was able to conclude that the radar retrieved normalised gamma distribution parameters, $N_w$ and $D_0$, are broadly "consistent with disdrometer measurements". However, this paper conceded that the shape parameter $\mu$ proved difficult to accurately predict.

Since the introduction of the normalised gamma distribution, several studies have demonstrated that it is possible to retrieve the parameters of the gamma model ($N_w$, $D_0$, $\mu$) from polarimetric radar measurements of $Z_h$, $Z_{dr}$, and $K_{dp}$. For instance, Gorgucci et al. (2001; 2002) developed the "β – method" for C-band radars which was then extended to S-band radars by Bringi et al. (2002). The "β – method" improved on earlier rainfall algorithms that were derived with a priori assumptions, namely, the equilibrium shape–size relationship. Consequently, the "β – method" exhibited immunity to variability in shape–size relation. Further, the "β – method" performed competently, estimating the drop median diameter ($D_0$) to an accuracy of 10%, and the (logarithmic-scale) equivalent intercept parameter ($N_w$) to an accuracy of 6%. However, $\mu$ proved difficult to estimate (Gorgucci et al., 2001; 2002). As an alternative to the "β – method", Zhang et al. (2001) proposed the "constrained-gamma (CG) method" which assumed that $N_w$, $D_0$, and $\mu$ are not mutually independent. However, Brandes et al. (2004) determined that the "β – method" was more consistent with empirical RSD observations that the "CG method" and was less sensitive to specific differential phase-shift ($K_{dp}$) errors.

More recent studies by Vulpiani et al. (2006) attempted to retrieve the gamma model parameters using neural networks (NN). The universal approximation theorem proves that NNs with a single hidden layer and a continuous and arbitrary bounded activation function can approximate any continuous function (Cybenko, 1989; Hornik et al., 1989). Hence, NNs are capable of approximating strongly nonlinear functions, including those that describe the relationship between radar observables and RSD parameters (Vulpiani et al., 2006). Thus, Vulpiani et al. (2006) was able to use NNs to estimate the parameters of the normalised gamma model from $Z_h$, $Z_{dr}$, and $K_{dp}$. This NN methodology was an improvement on previous approaches described in the literature and performed 'fairly well even for low specific differential phase-shift [$K_{dp}$] values' (Vulpiani et al., 2006). However, this research does not investigate whether alternative radar products could be used to estimate the normalised gamma model, nor does it experiment with a range of NN structures.

Similarly, another study by Pasarescu (2020) also developed a NN-based model to retrieve RSD parameters from polarimetric weather radar measurements. Her model also only used $Z_h$, $Z_{dr}$, and $K_{dp}$ as input features, and although she experimented with various NN architectures, she also only conducted her analysis with one hidden-layer NN with 22 neurons in the hidden layer.

An alternative RSD parameter retrieval technique is to use decision trees. For example, Conrick et al. (2020) used an ensemble of decision trees (i.e., a random forest) to retrieve the median volume diameter $D_0$, liquid water content (LWC), and rain rate from (S-band) polarimetric weather radar measurements ($Z_h$, $Z_{dr}$). The researchers also calculated $N_w$ as a function of LWC and $D_0$. The model was tested with a Parsivel² disdrometer dataset and was found to be "more accurate and unbiased than nonlinear regression retrievals" with "some improvements in RMSE exceeding 50%". Further, the model was "more conducive to obtaining DSDs that are physically accurate". However, this research did not investigate alternative DT structures and input parameters, nor compare the performance against a range of different analytical and machine learning models. Thus, despite significant advancements in RSD retrieval, there are several critical research gaps that need to be addressed:

1. How does the performance of Gorgucci's (2002) RSD retrieval method compare to the alternative machine learning (ML) approaches using the same data set? Without this research, it is not possible to accurately compare the efficacy of different methods for RSD retrieval.
2. How does varying the ML model's structure alter performance? Investigating a range of model structures could lead to more accurate ML models.
3. Finally, how does using different input sets affect the accuracy of RSD parameter estimates?

To answer these research questions, the paper has the following objectives. Obtain disdrometer data, filter anomalous and irrelevant results, and validate using rain gauge data. Then, find the measured, then gamma model of the RSD, and calculate the radar parameters. Finally, iteratively train and test the various models using the radar parameters as inputs, ensuring to experiment with a variety of input sets and ML structures.

## 3. DATA & METHODS

The data used for this paper was collected between the 22$^{nd}$ of September 2020 and the 11$^{th}$ of May 2022 from four collocated rain gauges located at Sheepdrove Farm in the south of England (see Figure 1). These gauges included one OTT Parsivel² optical disdrometer (see Figure 2), one OTT Pluvio² weighing bucket rain gauge (see Figure 3), and two tipping bucket rain gauges with buckets of 0.1mm and 0.2mm (see Figure 4).

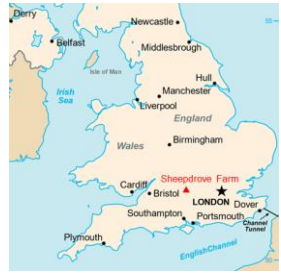

**Figure 1: Location of Sheepdrove Farm on map of England**

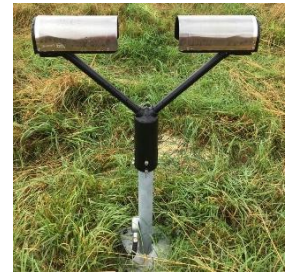

**Figure 2: OTT Parsivel² optical disdrometer rain gauge**

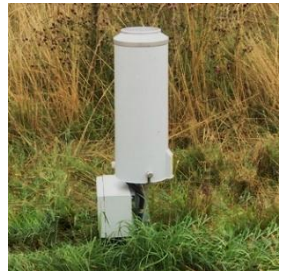

**Figure 3: OTT Pluvio² weighing bucket rain gauge**

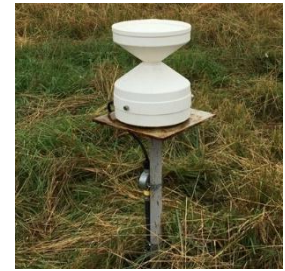

**Figure 4: Tipping bucket (0.2mm) rain gauge**

TBRGs are widely used instruments for measuring rainfall due to their simplicity and comparatively low cost (Strangeways, 2000). They consisted of a funnel-shaped collector that directs precipitation into one of a pair of buckets that are balanced bistably on a horizontal axis (American Meteorological Society, 2012). When a pre-determined amount of precipitation accumulates, the bucket tips, emptying the water and triggering a switch that sends a signal to a recording device (Nystuen et al., 1996). TBRGs tend to underpredict the rain rate during low-intensity rainfall events. This is partly because they require a critical quantity of water to accumulate in the bucket before tipping (Ciach, 2003). Their measurements are also impacted by both evaporation of precipitation that has landed in the gauge and wind disturbing the precipitation fall trajectory (Nešpor and Sevruk, 1999). Finally, TBRGs also typically underpredict the rain rate during intense rainfall events (Ciach, 2003) because the bucket cannot tip fast enough to capture all the precipitation (Luyckx and Berlamont, 2001). In contrast, weighing gauges, such as the OTT Pluvio², are less susceptible to errors associated with high-intensity rainfall events, but are more expensive and require greater maintenance (Duchon and Biddle, 2010). Weighing bucket rain gauges work by continuously weighing the precipitation that accumulates in the collection container (Nemeth, 2008).

Optical disdrometers (e.g., the Parsivel²) measure individual raindrops (Löffler-Mang and Joss, 2000) so typically offer superior accuracy compared to TBRGs during low and high intensity rainfall. However, unlike TBRGs and Pluvio weighing gauges, the Parsivel[2] disdrometer does not directly provide values for the rainfall rate. Instead it measures the size and velocity of individual raindrops using a narrow laser beam directed at an optical sensor (Tokay et al., 2014). As precipitation falls through the instrument's sampling area, it disrupts the light beam, and the resulting attenuation in light intensity causes the voltage output from the photodiode to reduce. This is subsequently used to automatically determine the drop size and velocity, allowing for the estimation of rainfall rate ($R_{disd}$) and hydrometeor type (Tokay et al., 2014). The drop sizes are provided in the form of hydrometeor diameters broken down into 32 unevenly spaced size classes ($n_i$) between 0.062mm and 24.5mm.

The Pluvio[2] and the TBRGs rain rates ($R_{pluvio}$ and $R_{TBRG}$ respectively) will be used to validate the data provided by the disdrometer. The accumulation interval time for all devices is one-minute, consistent with previous studies including: Tokay et al. (2003), Bringi et al. (2011), Adirosi et al. (2014), and Pasarescu (2020).

### 3.1 Filtering and Validation of Rain Gauge Data

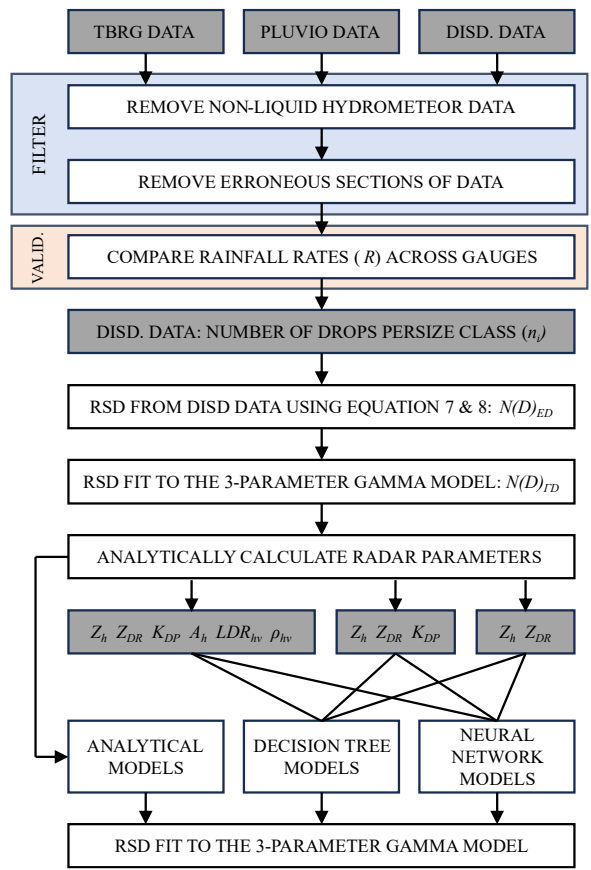

**Figure 5: Processes followed to obtain RSD and radar parameters from disdrometer data**

The present study uses data from the ground-based rain gauges to produce a set of analytically derived radar parameters. These radar parameters are then used to fit models that can estimate the three parameters of the normalised gamma model of the RSD. However, in order to improve the general reliability of the models, it was imperative that the training data was of the highest possible quality. Consequently, a filtering and validation process was developed. Figure 5 outlines the method that was followed to obtain the feature sets (i.e., the analytically simulated radar parameters). Because the present study is limited to the analysis of rain, the first step in the filtering process was to remove all the non-liquid hydrometeors. This was achieved using precipitation classification data provided by the OTT Parsivel[2] disdrometer. Data across all rain gauges were omitted for the period of time that the precipitation type was non-liquid (e.g., snow, hail, etc.).

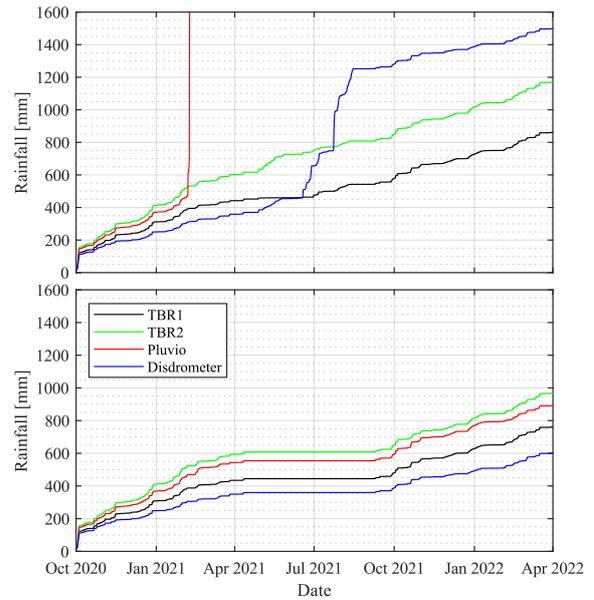

**Figure 6: Rainfall accumulation before (top) and after (bottom) filtering.**

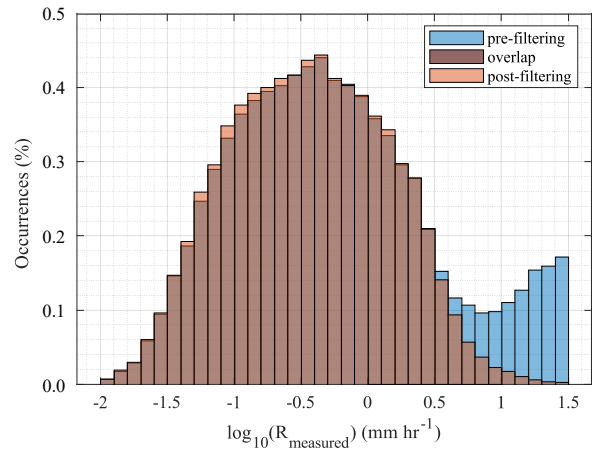

**Figure 7: Disdrometer rainfall histogram: before (blue/below) and after (orange/above) filtering.**

By plotting the accumulation of rainfall over the data gathering interval (approximately 597 days), it was possible to identify potentially erroneous sections of data – see Figure 6 (top). For instance, the Pluvio gauge recorded an extreme increase in rainfall (on approximately 06/02/2021) that was not corroborated by the other gauges. This data spike was likely due to a non-precipitation entity falling into the gauge – e.g., a small animal. To safeguard the reliability of the data, all data from all gauges was removed for the duration of the increase. The TBR1 data (corresponding to the accumulation of $R_{TBRG\ 0.1mm}$) failed to report data from roughly 30/04/2021 to 29/06/2021. It is probable that the tipping mechanism was prevented from rocking due to some mechanical block. This data was therefore omitted. Finally, for roughly two months (i.e., 16/06/2021 to 14/08/2021), the disdrometer recorded unexpectedly elevated precipitation accumulation, equivalent to the typical annual rainfall for the region. During this period, anomalously high raindrop diameters were also recorded. Following a maintenance visit, the disdrometer started returning more credible results. Thus, the most plausible cause of this erroneous period was vegetation growing in front of the laser sensor and therefore measuring large raindrops with negligible fall velocity. Consequently, as with the other errors identified, all the data from all the sensors was omitted for the erroneous time period. The final accumulation data is graphed in Figure 6 (bottom). Following this process, the gauge accumulations clearly track more closely. Figure 7 shows a histogram of the measured disdrometer rainfall before filtering (blue) and after (orange) – note the decrease in unexpectedly high rain rates.

In order to benchmark the quality of the disdrometer data, it was necessary to know the 'true' rainfall rate. However, all empirical measurement techniques are subject to errors. For instance, TBRGs exhibit systematic errors including those associated with wind and both heavy and light rainfall (e.g., evaporation, undercatchment, and splash-out) (Habib et al., 2001; Humphrey et al., 1997; Nemec, 1969). Consequently, an approximation of the 'true' rain rate would need to be constructed from the available data.

The data that is presently available comes from the collocated rain gauges and is provided in one-minute accumulation intervals. However, the discrete nature of TBRGs means that precipitation that has accumulated in the buckets may not have reached the critical threshold required to cause a tip within the one-minute interval. This means that the TBRG data would not reflect the correct rain rate for that accumulation interval. So, by accumulating the rain rate over larger sampling intervals (e.g., 30 minutes), these sampling related errors become minimally important (Ciach, 2003; Habib et al., 2001). Furthermore, Ciach (2003) demonstrated that averaging several rain gauges is an effective method of accurately estimating the true rain rate. Consequently, an estimate of the 'true' rain rate ($R_T$) was constructed by finding the arithmetical mean of the rain rate from the two TBRGs and the Pluvio gauge. This average was then accumulated over various time intervals ($T$). See Equation 1.

$$R_T = R_{TBRG\ 0.1mm,T} + R_{TBRG\ 0.2mm,T} + R_{Pluvio,T} \qquad 1$$

Once the 'true' rain rate was established, it was possible to derive insights into the quality characteristics of the disdrometer data. Equation 2 expresses the relative error ($e_T$) of the disdrometer data against the 'true' rain rate.

$$e_T = \frac{R_{disdrometer,T} - R_T}{R_T} \qquad 2$$

Figure 8 illustrates the relative error plotted against the 'true' rain rate. This qualitatively indicates that the relative error dispersion is high for low rainfall intensities. However, consistent with Ciach's (2003) results, these errors decrease relatively quickly as rainfall intensity increases.

To draw more quantitative conclusions about the data, a nonparametric kernel regression estimator was used. This method evaluated the standard deviations of the relative errors as a function of the 'true' rainfall intensities. This approach is consistent with the method adopted by Ciach (2003). Equation 3 provides Nadaraya–Watson's (N-W) kernel regression estimator for the square of the standard error ($\sigma_i^2$) for a particular disdrometer rainfall value $R_{T_i}$ at a fixed accumulation time ($T$) and it is given by:

$$\sigma_i^2(T, R_T) = \frac{\sum_i K\left(\frac{R_T - R_{T_i}}{h}\right) e_{T_i}^2}{\sum_i K\left(\frac{R_T - R_{T_i}}{h}\right)} \qquad 3$$

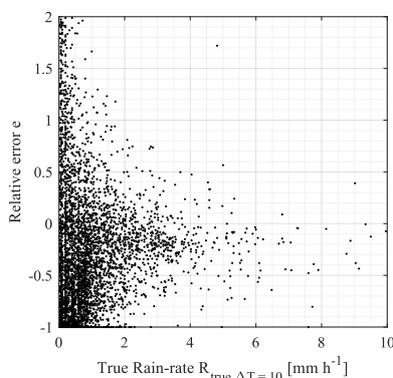

**Figure 8: Scattergram of the relative errors of the disdrometer against the 'true' rain rate at a 10-minute time**

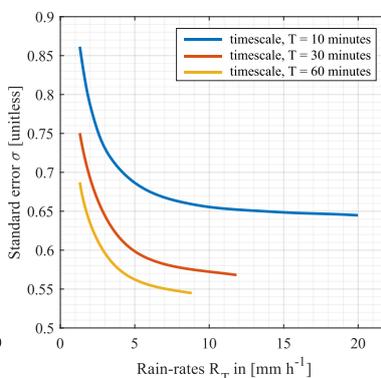

**Figure 9: Standard deviation of the disdrometer's relative errors of the against the 'true' rain rate at 10, 30, and 60-minutes.**

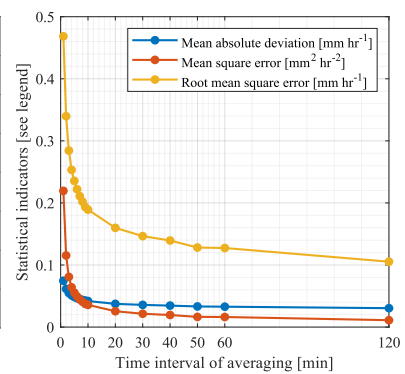

**Figure 10: MAE, MSE, and RMSE of $R_{disd}$ for different accumulation time intervals.**

Where $R_T$ is the full set of disdrometer rainfall data, $e_{T_i}^2$ is the second statistical moment of the relative error for a particular 'true' rainfall value, $h$ is the estimation bandwidth, and $K(\cdot)$ is the kernel. For this analysis, in line with Ciach's (2003) approach, the Epanechnikov kernel was used – see Equation 4.

$$K(x) = \begin{cases} 0.75(1 - x^2), & x \in [-h, h] \\ 0, & x \notin [-h, h] \end{cases} \qquad 4$$

In N-W regression, using a higher bandwidth results in a smoother fit at the expense of fine details. Ciach (2003) opted to use a variable bandwidth equivalent to 20% of the specific disdrometer rainfall value: in contrast, the present analysis (see Figure 9) uses a bandwidth equivalent to 100% of the specific disdrometer rainfall value.

In line with expectations, the analysis depicted in Figure 9 concluded that the relative error decreases as rain rate increases. Further, it can be concluded that accumulating over larger time intervals reduces the overall error. However, it is necessary to extend the present analysis to determine how the changing accumulation interval impacts the error. Thus, in order to quantify the difference between the disdrometer rain rate and the true rain rate, three performance indicators will be utilised: the mean absolute deviation (Equation 5), the root mean square error (Equation 6), and the mean square error (Equation 6).

$$MAD = \frac{\sum_{i=1}^{N}(R_T - R_{disd,T})}{N} \qquad 5$$

$$RMSE = \sqrt{MSE} = \sqrt{\frac{\sum_{i=1}^{N}(R_T - R_{disd,T})^2}{N}} \qquad 6$$

Where $N$ is the number of individual accumulated rain rates recorded (for $R_{disd,T}$ and $R_T$ see Equation 1 and 2).

As previously shown by the N-W regression, there is a tendency for the error to decrease as the accumulation interval is increased. Figure 10 further expands on this relationship. It can be seen that there is a significant decrease in the error as the accumulation is initially increased from one-minute, however, the returns become increasingly marginal. Thus, for the purposes of validating the accuracy of the disdrometer rainfall data, the interval of 60 minutes is selected as it significantly decreases error but retains a large set of data.

Finally, Figure 11 shows the scatter between all the rain rates, the RMSE, and linear regression line. From these results it is apparent that for a 60-minute accumulation interval, the disdrometer rain rate has an RMSE equal to 0.127 [mm hr$^{-1}$] and that it tends to underpredict the rain rate. However, for the purposes of the validation of the disdrometer data, it is evident that there is a close relationship between the 'true' value of rain rate and the disdrometer's measurements (RMSE = 0.127 [mm hr$^{-1}$]). Consequently, the rest of the analysis will proceed with the assumption that the disdrometer's data is sufficiently accurate to train the models.

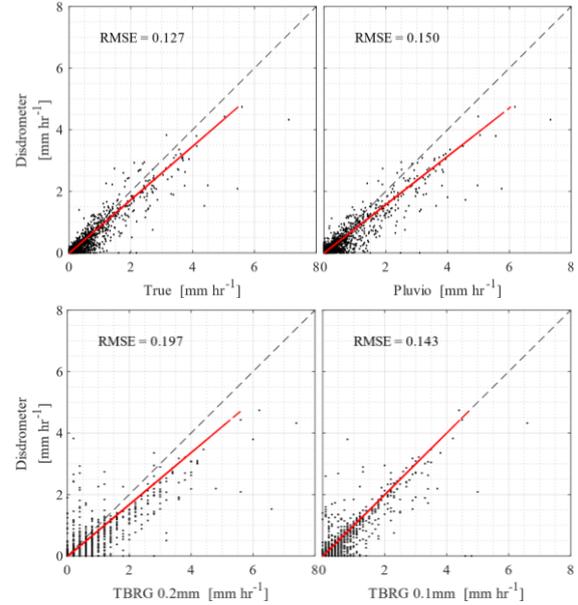

**Figure 11: Composite scatterplot of the disdrometer rain rates against the other rain rates for a 60-minute accumulation interval. The linear regression is shown as a red line.**

### 3.2 Obtaining the Parameters of the Three-Parameter Normalised Gamma Model

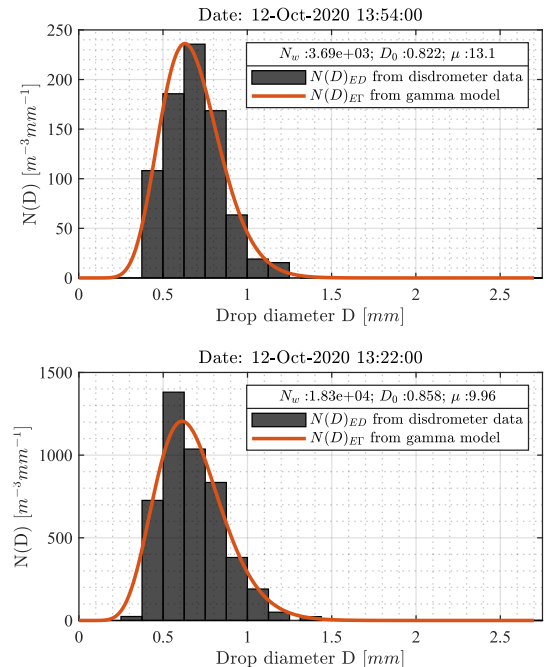

**Figure 12: Histogram of RSD from disdrometer data and from the three-parameter gamma model**

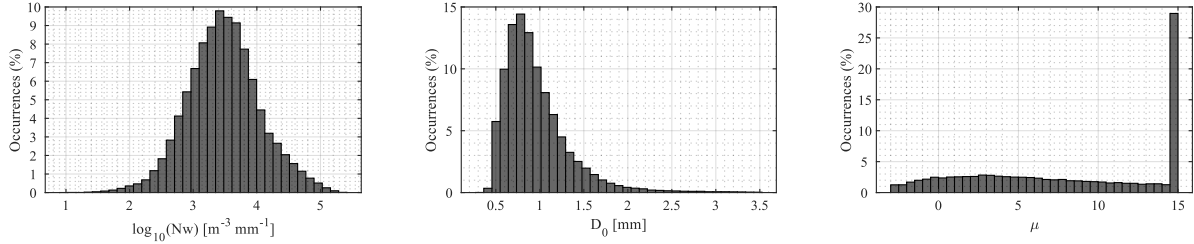

**Figure 13: Histogram of the fitted normalised gamma distribution parameters ($N_w, D_0, \mu$)**

Following the filtering and validation process, it can be assumed that the remaining disdrometer data was sufficiently accurate to continue the analysis. Critically, this data included the hydrometeor diameters broken down into 32 size classes ($n_i$) at one-minute intervals. To compare disdrometer rainfall measurements with TBRGs, a longer time interval was needed to compensate for the errors in TBRG measurements. However, when retrieving the RSD, it is more appropriate to use the original one-minute accumulation interval. This is because this temporal resolution is more consistent with the instantaneous radar rainfall observations as it is brief enough to capture the everchanging rainfall variability.

In order to retrieve the parameters of the three-parameter normalised gamma model (hereon, 3PNG model), it was first necessary to formulate an estimate of the RSD at the discrete one-minute instant ($t$) from the data provided by the disdrometer using Equation 7 (Montopoli et al., 2008):

$$RSD, N_{ED}(D_i, t) = \frac{n_i(t)}{A \cdot \Delta t \cdot v_i \cdot \Delta D_i} \quad [m^{-3} mm^{-1}] \quad 7$$

Here, $N_{ED}(D_i, t)$ represents the number of drops present in the size class ($n_i$), the subscript $ED$ denotes that the result is estimated from disdrometer data, $D_i$ denotes the diameter width of the size class (in mm), $\Delta t$ represents the time interval (one-minute), $A$ denotes the disdrometer measuring area, and $v_i$ denotes the rain fall speed. However, rain fall speed varies depending on the size of the droplet. For the present analysis, the velocity of a certain raindrop is given in Equation 8 (Atlas et al., 1973):

$$v_i = 9.65 - e^{-0.6 D_i} \quad [m\ s^{-1}] \quad 8$$

As outlined in Section 1, there is a consensus in the literature to represent the RSD using the normalised gamma model. The present analysis shall use the form provided by Bringi et al. (2003) – see Equation 9:

$$RSD, N_{E\Gamma}(D) = N_w \cdot f(\mu) \cdot \left(\frac{D}{D_m}\right)^\mu \cdot e^{\frac{-(4+\mu) \cdot D}{D_m}} \quad [m^{-3} mm^{-1}] \quad 9$$

Here, the subscript $E\Gamma$ denotes that the RSD has been estimated using the 3PNG model, $N_w$ is the normalised drop concentration parameter (in $mm^{-1}m^{-3}$), $D_m$ is the mass-weighted mean drop diameter parameter (in mm), $\mu$ is the gamma shape parameter (unitless), and $f(\mu)$ is given in Equation 10:

$$f(\mu) = \frac{6(4+\mu)^{\mu+4}}{4^4 \Gamma(\mu+4)} \quad 10$$

Here, $\Gamma$ is the gamma function. Finally, for this analysis, the median volume drop diameter ($D_0$) is used instead of the $D_m$ parameter, Ulbrich (1983) provides the conversion described by Equation 11:

$$\frac{D_0}{D_m} = \frac{3.67 + \mu}{4 + \mu} \quad 11$$

Given these relationships, it was possible to iteratively fit the 3PNG model of the RSD ($N_{E\Gamma}$) to the RSD provided by the disdrometer data ($N_{ED}$). Thus, a set of parameters ($N_w, D_0, \mu$) was obtained that could be used to represent the 3PNG model RSD ($N_{E\Gamma}$) for each (non-omitted) minute of data over the whole recording interval (22nd of September 2020 to the 11th of May 2022). Figure 12 shows the fit of a 3PNG model RSD ($N_{E\Gamma}$) (red line) to a histogram of the RSD provided by the disdrometer data ($N_{ED}$) for an arbitrary selection of data. The histogram in Figure 13 provides a summary of all the parameters.

### 3.3 Obtaining Radar Parameters Analytically from the Normalised Gamma Distribution RSD

So far, the data from the disdrometer has been filtered, validated, and used to produce an RSD ($N_{ED}$). Then, that RSD has been used to fit the three-parameter normalised gamma model (3PNG model) of the RSD ($N_{E\Gamma}$). In this section, analytical equations will be used to derive estimates of the radar parameters using the 3PNG model RSD as an input. Subsequently (in Section 3.4), these radar parameters will be used to fit models that estimate the parameter 3PNG model ($\widehat{N}_w, \widehat{D}_0, \hat{\mu}$) (see Figure 5 for method structure outline). The literature provides various established analytical relationships required to derive radar parameters from the RSD. For instance, Bringi and Chandrasekar (2001) provide the following relationships (Equation 12, 13, and 14) used to evaluate the horizontal or vertical reflectivity ($Z_{h,v}$), the differential reflectivity ($Z_{dr}$), and the specific differential phase shift ($K_{dp}$).

$$Z_{h,v} = \frac{4\lambda^4}{\pi^4 |K_w|^2} \int_{D_{min}}^{D_{max}} |S_{hh,vv}(D)|^2 N(D) dD \quad [mm^6 m^{-3}] \quad 12$$

$$Z_{dr} = 10 \log_{10} \frac{Z_h}{Z_v} \quad [dB] \quad 13$$

$$K_{dp} = 10^{-3} \frac{180}{\pi} \lambda \Re \left\{ \int_{D_{min}}^{D_{max}} [f_{hh}(D) - f_{vv}(D)] N(D) dD \right\} \quad 14$$

Here, $D$ is the raindrop equivalent volume diameter, $N(D)$ is the raindrop size distribution, $D_{max}$ and $D_{min}$ is the maximum and minimum diameter in a size distribution, $\lambda$ is the radar wavelength, $S(D)$ the complex scattering matrix of a raindrop, $S_{hh,vv}(D)$ is the backscattering copolar components of S, $\Re$ returns the real component of a complex number, and $f_{hh}$ and $f_{vv}$ are the forward scattering amplitudes at horizontal and vertical polarisations.

Further, Bringi and Chandrasekar (2001) also provide equations for the linear depolarisation ratio ($LDR_{vh}$), and the correlation coefficient between pairs of $Z_h$ and $Z_v$ measurements ($\rho_{hv}$). The linear depolarisation ratio is a measure of how much of the backscattered electromagnetic signals have changed polarisation after interacting with the target precipitation. The linear depolarisation ratio is expressed as a dimensionless quantity between nought and unity, where nought indicates that the backscattered radiation is entirely unpolarised, and unity indicates that the backscatter is completely polarised. Generally, more anisotropic targets (e.g., ice crystals) have higher linear depolarisation ratios (Bringi and Chandrasekar, 2001).

The correlation coefficient between pairs of $Z_h$ and $Z_v$ measurements ($\rho_{hv}$) for rain tends to be close to unity. However, the heavier the rainfall, the more oblate the raindrops become. Therefore, $Z_h$ becomes larger than $Z_v$ and so the correlation decreases. For other non-liquid hydrometeors (e.g., snow, ice, melting snow) the correlation decreases more rapidly (Bringi and Chandrasekar, 2001). The final parameter is the specific attenuation ($A_H$) which has been used in the form provided by Gu et al. (2011). Radar signal attenuation usually occurs during heavy rain and affects radar measurements with higher frequencies (e.g., >5GHz).

## 3.4 Models Used to Estimate the Three Parameters of the Gamma Model from the Radar Parameters

The present section introduces the models that are trained using the simulated radar parameters established in Section 3.3 ($Z_h, Z_{dr}, K_{dp}, A_H, LDR_{vh}, \rho_{hv}$) to estimate the 3PNG model parameters ($\widehat{N}_w, \widehat{D}_0, \hat{\mu}$). These models included analytical equations (ANs), neural networks (NNs), and decision trees (DTs). Consistent with Vulpiani et al. (2009; 2006), a random 70% of the simulated radar data will be used to train the models, and the remaining 30% will be used to test the models. The training and testing process will be repeated 100 times in order to account for model equifinality and to enable cross-validation of the model structures (Beven, 2006).

Unlike analytical equations, where the choice of input parameters is determined by the specific equation, machine learning models are capable of using a broad range of input parameters (or "features"). Consequently, three separate "feature sets" were proposed to account for the varying availability and reliability of radar products in real world applications. The sets are as follows:

- Feature set 1: $\{Z_h, Z_{dr}, K_{dp}, A_H, LDR_{vh}, \rho_{hv}\}$
- Feature set 2: $\{Z_h, Z_{dr}, K_{dp}\}$
- Feature set 3: $\{Z_h, Z_{dr}\}$

The horizontally polarised radar reflectivity factor ($Z_h$) and the differential reflectivity ($Z_{dr}$) are often used in RSD parameter retrieval (Gorgucci et al., 2002). Further, Vulpiani et al. (2006) noted that the specific differential phase shift ($K_{dp}$) is a plausible predictor for the RSD, however, its computation is often noisy so may have a deleterious impact on the results. Thus, consistent with Vulpiani et al. (2006), the feature set 2 and 3 represent options for when $K_{dp}$ is, and is not, reliable. Finally, feature set 1 extends the analysis to evaluate how a broader range of features impacts the results.

### 3.4.1 Analytical Equations

Given certain weather radar parameters ($Z_h, Z_{dr}, K_{dp}$), Gorgucci et al. (2002) demonstrated that it is possible to estimate the parameters of the 3PNG model ($\widehat{N}_w, \widehat{D}_0, \hat{\mu}$). For instance, the median volume drop diameter ($D_0$) can be obtained from Equations 15, 16 (Gorgucci et al., 2002), and 17 (Bringi et al., 2002; Gorgucci et al., 2001):

- $\widehat{D}_0 = AN_1(Z_h, Z_{dr}) = a_1 Z_h^{b_1} (\xi_{dr})^{c_1}$   [mm]          15

- $\widehat{D}_0 = AN_2(K_{dp}, Z_{dr}) = a_2 K_{dp}^{b_2} (\xi_{dr})^{c_2}$   [mm]          16

- $\widehat{D}_0 = AN_3(Z_{dr}) = a_3 Z_{dr}^{b_3}$   [mm]          17

Here, the coefficients $a_i, b_i, c_i$ are determined by curve-fitting the equations to test data, $\xi_{dr} = 10^{0.1 Z_{dr}}$ is the differential reflectivity in the linear scale, $Z_{dr}$ is the differential reflectivity in decibels, and $Z_h$ is the reflectivity factor at horizontal polarisation ($mm^6 m^{-3}$). The intercept parameter ($N_w$) of the 3PNG model is provided by Equations 18 and 19 (Gorgucci et al., 2002):

- $\log_{10} \widehat{N}_w = AN_4(Z_h, Z_{dr}) = a_4 Z_h^{b_4} (\xi_{dr})^{c_4}$          18

- $\log_{10} \widehat{N}_w = AN_5(K_{dp}, Z_{dr}) = a_5 K_{dp}^{b_5} (\xi_{dr})^{c_5}$          19

Finally, the shape parameter ($\mu$) is provided in Equation *20* (Gorgucci et al., 2002):

- $\hat{\mu} = AN_6(\widehat{D}_0, Z_{dr}) = \frac{a_6 (\widehat{D}_0)^{b_6}}{(\xi_{dr} - 1)} - c_6 (\xi_{dr})^{d_6}$          *20*

Here, $\widehat{D}_0$ is the estimated values of median volume drop diameter obtained in Equations 15, 16, and 17. Thus, there are three versions of Equation 20 depending on the specific median volume drop diameter estimate used.

### 3.4.2   Neural Networks & Decision Trees

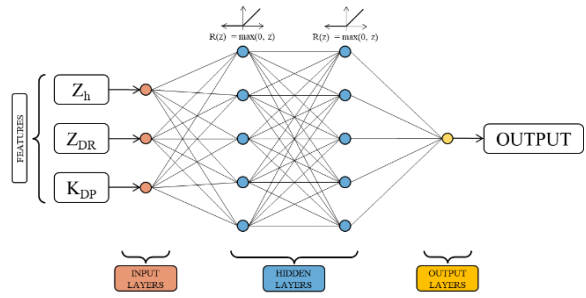

**Figure 14: Structure of a two-hidden-layer neural network regression model with three features (inputs). The ReLU activation function is shown.**

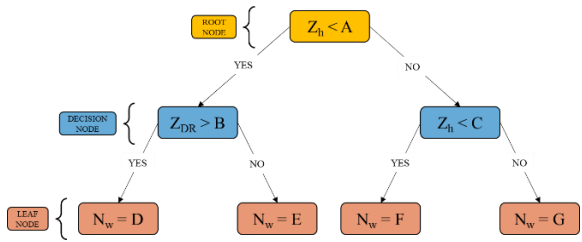

**Figure 15: Structure of a four-leaf decision tree. Root node, decision nodes, and lead nodes are labelled, and colour coded.**

Figure 13 depicts a particular regression neural network (NN) with two-hidden layers, three-input parameters (called features), and a single output node. However, in this study, a more extensive range of NN structures will be used; specifically, there will be three options for the number of hidden layers {1,2,3}, five options for the number of neurons per hidden layer {10,20,30,40,50}, and three sets of features. This range of options corresponds to 45 unique NN structures. Consequently, unlike research from Vulpiani et al. (2009; 2006) which uses the sigmoid activation function, this study will require a more computationally efficient (i.e., faster) Rectified Linear Unit (ReLU) activation function (Goodfellow et al., 2016). Each NN structure was fit to each of the three parameters of the gamma model 100 times: this corresponds to 13500 NNs. Equation 21 describes a NN:

$$\widehat{D}_0 = NN_{L,N}(Feature\ set) \qquad 21$$

Here, $L$ is the number of layers, $N$ is the number of neurons, the left-hand side of the equation denotes the output, and the argument of the function is the feature set used to train the NN model.

Like the NN, the decision trees (DT) will have a variety of structures. Specifically, they will have a variable number of leaf nodes {4,12,36}. An arbitrary DT is shown in Figure 15. The decision tree models will use the same feature sets as the NN models. Equation 22 describes a DT:

$$\widehat{D}_0 = DT_L(Feature\ set) \qquad 22$$

Here, $L$ is the number of leaf nodes.

## 4.   RESULTS

Thus far, the 3PNG model of the RSD ($N_{E\Gamma}$) has been obtained and used to analytically calculate the simulated values of the radar parameters. These radar parameters were then used to train a large number of analytical, neural network, and decision tree models 100 times, each from a random 70% of the simulated radar parameters. The remaining 30% of the data was reserved for testing the models. In total, 17000 unique models were produced.

Figure 16 shows a scatter plot of the 3PNG parameters obtained from the disdrometer data ($N_w, D_0, \mu$) against the 3PNG parameters ($\widehat{N}_w, \widehat{D}_0, \hat{\mu}$) obtained from a NN model with 1 layer, 10 neurons, and using set 1 as the input features – i.e., $NN_{1,10}(Z_h, Z_{dr}, K_{dp}, A_H, LDR_{vh}, \rho_{hv})$. Critically, these output parameters ($\widehat{N}_w, \widehat{D}_0, \hat{\mu}$) were from the testing stage – not the training stage. The feature set provided to the model during the testing stage was the 30% of data that had not been used to train the model. As expected, if the 70% of features that trained the model were fed back into the model, the output parameters were generally closer to the true parameters.

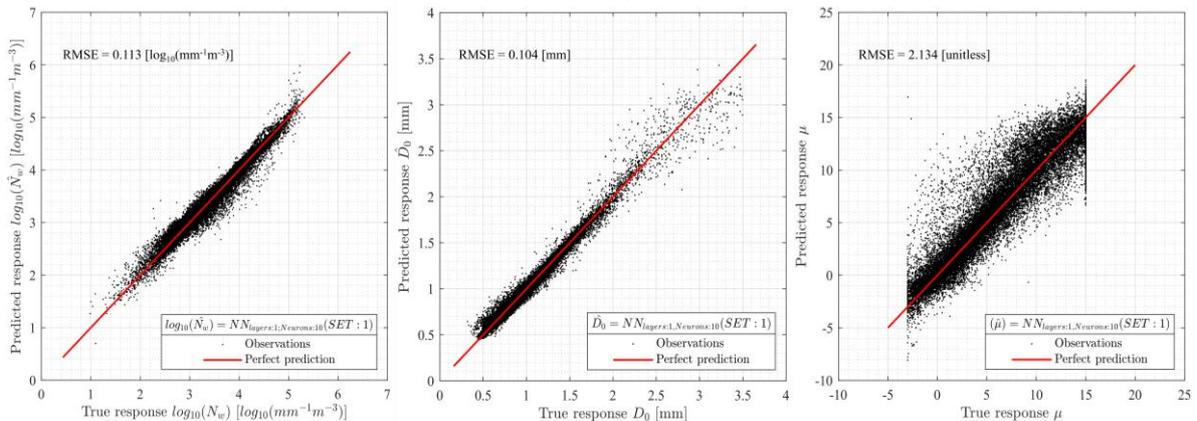

**Figure 16: Scatterplot of $N_w$, $D_0$, and $\mu$ against their corresponding predicted values. The structure of the model was $NN_{L1,N10}$ and the feature set was set 1 $\{Z_h, Z_{dr}, K_{dp}, A_H, LDR_{vh}, \rho_{hv}\}$.**

The results in Figure 16 appear to show that the model was reasonably effective at approximating the 'true' values of the normalised drop concentration parameter ($N_w$) and the median volume drop diameter ($D_0$), but it was less effective at approximating the gamma shape parameter ($\mu$). For the (log base 10) $N_w$ scatter, the model seems to have strong predictive utility. However, closer inspection reveals that the model struggles to predict higher $N_w$ values (i.e., greater than 5). This is shown in the scatter as a tendency of the data to trend upwards, indicating that the predicted values ($\widehat{N}_w$) are greater than the 'true' $N_w$ values. Despite this, the vast majority of the roughly 33962 data points appear close to the perfect prediction line.

Similarly, for the median volume drop diameter ($D_0$), the model appears to be skilful at predicting the lower $D_0$ values, but less capable when presented with higher $D_0$ values (>2mm). In this case, although the scatter is distributed either side of the perfect prediction line, the scatter appears to trend downwards. This indicates that the model is systematically underpredicting high $D_0$ values. As before, it is worth noting that this represents a minority of the 33962 data points. Consistent with the findings of Bringi et al. (2003) and Vulpiani et al. (2006), the gamma shape parameter ($\mu$) proved more challenging to accurately predict using this particular model. The 'true' values of $\mu$ were confined to between -3 and 15. This was because values outside that range are unrealistic in natural DSDs (Bringi et al., 2003), however, the predicted values do not reflect this boundary.

Figure 16 shows just 1 of the 17000 models that were trained. In order to appropriately compare the extensive range of models, a more quantitative and rigorous analysis was needed. This analysis is presented in Figure 17. As before, these results are for only the testing (not training) phase.

As shown in Figure 17, the analytical models ($AN$) produced results of consistent accuracy, this is indicated by the narrow range of RMSE values. The results show that $AN_4(Z_h, Z_{dr})$ predicted $N_w$ more accurately than $AN_5(K_{dp}, Z_{dr})$, thus suggesting that $Z_{dr}$ is a more useful radar parameter than $K_{dp}$ when predicting $N_w$ for this particular analytical equation. As for $D_0$, the results showed that $AN_1(Z_h, Z_{dr})$ and $AN_2(K_{dp}, Z_{dr})$ had similar results, but both were outperformed by $AN_3(Z_{dr})$. Finally, for $\mu$, the $AN_{6A}(\widehat{D}_0, Z_{dr})$ and $AN_{6B}(\widehat{D}_0, Z_{dr})$ models performed roughly the same. This is, perhaps, not surprising as they used the similarly performant $AN_1$ and $AN_2$'s output as their $\widehat{D}_0$ inputs (respectively). Interestingly, the $AN_{6C}(\widehat{D}_0, Z_{dr})$, which used the more skilful $AN_3$ model's value of $\widehat{D}_0$, underperformed when compared to $AN_{6A}$ and $AN_{6B}$.

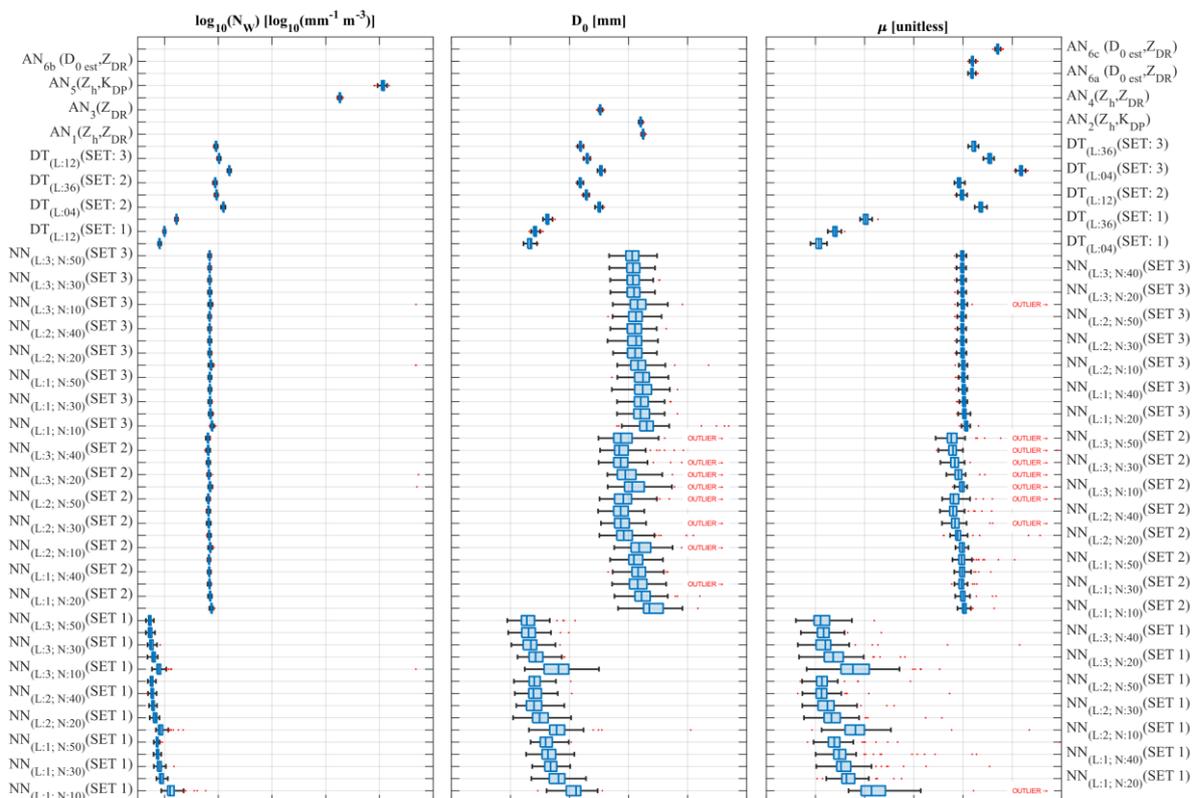

**Figure 17: Serial box plot of the RMSE of the testing process for each of the model structures and feature sets. Each box plot is derived from 100 data points, representing the iterations of the training and testing process. Each row corresponds to a model, labels are shown on the right and left.**

As can be seen in Figure 17, when predicting $N_w$, $D_0$, and $\mu$, the NN models that used feature set 1 (SET1) exhibited consistently lower median RMSEs than the NNs that used SET2 or SET3. Furthermore, for NNs that used SET1, the median RMSE tended to decrease as the number of both layers and neurons increased. This indicated that models with higher structural complexity were more performant when provided all the input features.

As for the NNs that used SET2, the RMSE range for $N_w$ was narrower than the results from SET1. Additionally, unlike for SET1, minimal $N_w$ prediction performance gains were realised when the model's structural complexity increased. For instance, $NN_{L:1,N:10}$ had a median RMSE of 0.184 and $NN_{L:3,N:50}$ had a median RMSE of 0.187. When predicting $D_0$ and $\mu$, the models that used SET2 returned less consistent RMSEs than those using SET1. This may have been due to the noise associated with $K_{DP}$. Further, increasing the number of layers and nodes of the SET2 NNs lowered the median RMSE for $D_0$ and $\mu$ predictions. However, more complex SET2 NNs produced a significant number of $D_0$ and $\mu$ RMSE outliers, possibly due to overfitting.

SET3 NNs performed similarly to SET2 NNs at predicting $N_w$ and had a higher degree of consistency than SET2 when predicting $D_0$ and $\mu$. SET3 NNs showed extremely minor improvement with increased complexity. Finally, despite the prediction of $N_w$ being fairly consistent across all feature sets and structures, some of the NN models with 2-3 layers and 10-20 neurons per layer occasionally exhibited an extremely high RMSE in comparison to the other RMSEs that model produced.

The decision trees (DT) that used SET1 had lower median RMSEs than those which used SET2 or SET3 for all three parameters (see Figure 17). Further, the performance DTs that used SET1 tended to decrease as the number of leaf nodes (i.e., structural complexity) was increased. For instance, when predicting $\mu$, $DT_{L:04}(SET1)$ returned a median RMSE of 1.535, and the more complex $DT_{L:36}(SET1)$ returned a median RMSE of 2.010. In contrast, DTs that used SET2 or SET3 to predict any of the parameters saw their results improve as the number of leaf nodes was increased from 4 to 36. The DT results were highly consistent and did not include extreme outliers. Further, the choice of input set appeared to have minimal impact on the spread of the RMSE results.

In summary, the machine learning models typically outperformed the analytical models. Specifically, ML models that used SET1 performed best. ML models that used SET2 also showed marginally lower median RMSEs than SET3, however, these models produced a much larger number of extreme outliers in NN models. Out of the ML models, the NN and DT models both showed promising utility. Broadly, the NN model results showed a lower consistency than the DT model results. However, no single model demonstrated consistently better performance at predicting any given RSD parameter. Similarly, varying the structure of the model resulted in remarkable improvements in model performance in some cases, and notable deteriorations in the predictive utility in other cases. Thus, it cannot be concluded that a single structure or model type is better. For instance, decision trees with 36 leaf nodes proved to be the most capable model at accurately obtaining the $D_0$ from SET3, but neural networks proved more capable at finding $N_w$ from SET1. Consistent with previous research by Bringi et al. (2003) and Vulpiani et al. (2006), $\mu$ proved more tricky to accurately predict than the $N_W$ and $D_0$.

## 5. DISCUSSION

In recent years, several studies have been conducted that develop models for the retrieval of RSD parameters using various techniques such as analytical relationships, neural networks, and random forests. However, no study has compared all these diverse approaches. Further, no study has so comprehensively evaluated the wide range of radar products and model structures for RSD retrieval, nor conducted such a rigorous analysis of model variability through repeated analysis.

For the purposes of this discussion, the most performant models from the analysis shown in Figure 17 have been selected to compare against those found in the literature (Table 1). We will neglect models that used feature set 2 (SET2) as $K_{dp}$ seem to only improve the performance at significant cost to the model's consistency.

|  | SET1 model | Q1 | Q2 | Q3 |
|---|---|---|---|---|
| $log_{10} N_w$ (mm$^{-1}$m$^{-3}$) | $NN_{L3,N50}$ | 0.070 | 0.072 | 0.074 |
| $D_0$ (mm) | $DT_{L4}$ | 0.065 | 0.067 | 0.068 |
| $\mu$ (unitless) | $DT_{L4}$ | 1.505 | 1.535 | 1.560 |
|  | SET3 model | Q1 | Q2 | Q3 |
| $log_{10} N_w$ (mm$^{-1}$m$^{-3}$) | $NN_{L1,N30}$ | 0.184 | 0.185 | 0.185 |
| $D_0$ (mm) | $DT_{L36}$ | 0.109 | 0.109 | 0.112 |
| $\mu$ (unitless) | $NN_{L2,N30}$ | 2.978 | 2.993 | 3.008 |

**Table 1: Quartile 1, 2, and 3 of the RMSE for the most performant models using feature sets (FS) 1 and 3.**

The neural network developed by Vulpiani et al. (2006) used measurements of $Z_h$, $Z_{dr}$, and $K_{dp}$ for retrieving the $D_0$ and $N_w$ parameters. Measurements of $Z_{dr}$ and the (previously estimated) $D_0$ parameter were also used to retrieve the shape parameter $\mu$. The model that the researchers used had a single hidden layer with six neurons for $D_0$ and $N_w$, and 12 neurons for $\mu$. In the paper, the researchers stated that their model had "shown an improvement with respect to other methodologies described in the literature to estimate the RSD parameters". Vulpiani's models had higher RMSEs for $D_0$ (0.29mm) and $N_w$ (0.28 mm$^{-1}$m$^{-3}$), but lower RMSEs for $\mu$ (1.27), than the corresponding models in Table 1. This paper's findings suggest that the neural network used by Vulpiani ($NN_{L1,N6\ or\ 12}$) was not the optimal structure. Additionally, Vulpiani used $K_{dp}$, which does not appear to improve performance, instead, the model could have benefitted from using an extended range of features, similar to those in SET1. Finally, the RMSEs indicate that the technique adopted by Vulpiani of predicting $\mu$ using $Z_{dr}$ and $D_0$ is potentially a superior approach.

Pasarescu (2020) also developed a NN-based model to retrieve RSD parameters from weather radar measurements ($Z_h$, $Z_{dr}$, and $K_{dp}$). She used a single-hidden-layer NN with 22 neurons in the hidden layer. Pasarescu's model achieved an RMSE of 0.09 (mm) for $D_0$, 0.26 (mm$^{-1}$m$^{-3}$) for $N_w$, and 1.57 for $\mu$. The results of this study suggest that Pasarescu's model outperforms Vulpiani's model in terms of the estimation of $D_0$, with an RMSE almost three times smaller. However, the RMSE for $\mu$ is larger than that of Vulpiani's model. Despite this, this paper shows that there is often significant variation in the RMSE when input parameters are iteratively randomised. Consequently, it cannot always be rigorously concluded that one model was better than another. Pasarescu's approach could have been improved by using SET1, or SET3 if SET1 was not available. Further, for prediction of (for instance) $D_0$, her approach of using NN models over decision trees (DT) appears to be less sound.

Conrick et al. (2020) used analytical methods and an ensemble of decision trees (i.e., a random forest) to retrieve the median volume diameter $D_0$ and other parameters from polarimetric weather radar measurements ($Z_h$, $Z_{dr}$). When retrieving $D_0$, the results from the analytical approach had an RMSE of 1.75 (mm), which decreased to 0.15 (mm) for the random forest. This paper supports Conrick's assertion that DTs show promising skilfulness at predicting the RSD, as both the most performant models for predicting $D_0$ were determined to be DTs.

A broader consideration when selecting the choice of models is the computational power available. For instance, complex NNs require significantly more time and data to train than the analytical regression models.

## 6. CONCLUSION

In conclusion, the machine learning approaches are usually more accurate than the analytical methods at obtaining the RSD parameters. However, no single machine learning model showed consistently more skill at predicting the parameters of the RSD. Instead, the choice of ML model and the way that model is structured should be contingent on the range of radar data obtainable and the particular parameter the model is finding. For instance, $DT_{L4}$ has shown considerable skill at predicting $D_0$ given SET1 data. The data used to train the model was observed to have a significant impact on the ability of the model to accurately predict the RSD parameters. Typically, SET1 performed the best. However, if SET1 data is not available, SET3 produces similar results to SET2, but with a greater level of consistency.

## 7. LIMITATIONS AND RECOMMENDATIONS OF FUTURE WORK

The machine learning approaches leveraged in this research assume that that the radar measurements are error free. However, it is well documented that measurements of the radar reflectivity parameters (including $Z_{dr}$ and $K_{dp}$) are noisy. Future work could introduce noise to the derived radar data to simulate this phenomenon. Further, the data used to train the models were from a specific location in the UK. This research could be extended by providing data from different climatic regimes. Thus, likely improving the generalisability. Finally, the literature would benefit from a similar study that tests a comparable range of structures and parameters, but this time, uses operational weather radar from an extensive and diverse range of locations to test the generality of different models.